\documentclass[sn-mathphys-num, iicol]{sn-jnl}

\usepackage{graphicx}%
\usepackage{multirow}%
\usepackage{amsmath,amssymb,amsfonts}%
\usepackage{amsthm}%
\usepackage{mathrsfs}%
\usepackage[title]{appendix}%
\usepackage{xcolor}%
\usepackage{textcomp}%
\usepackage{manyfoot}%
\usepackage{booktabs}%
\usepackage{algorithm}%
\usepackage{algorithmicx}%
\usepackage{algpseudocode}%
\usepackage{listings}%

\usepackage{lineno}

\usepackage[colorinlistoftodos]{todonotes}

\begin{document}

\title{Adiabatic lapse rate estimation using a van der Waals-type equation of state}



\author[1]{\fnm{J. C.} \sur{Hernández}}

\author[1]{\fnm{M. A.} \sur{Machucho}}

\author*[2]{\fnm{J. E.} \sur{Ramírez}}\email{jhony.ramirezcancino@viep.com.mx}

\affil[1]{Facultad de Ciencias F\'isico Matem\'aticas, Benem\'erita Universidad Aut\'onoma de Puebla,
Apartado Postal 165, 72000 Puebla, Puebla, Mexico}

\affil[2]{Centro de Agroecología, Instituto de Ciencias, Benemérita Universidad Autónoma de Puebla, Apartado Postal 165, 72000 Puebla, Puebla, M\'exico}


\abstract{
We revisit a family of temperature-dependent van der Waals-type equations of state (EOS) to improve the estimation of the adiabatic lapse rate in planetary atmospheres. These EOS generalize the classical van der Waals and Berthelot models by introducing a single parameter that modulates the temperature dependence of intermolecular interactions. We analyze their thermodynamic properties, including critical behavior, spinodal and coexistence curves, and entropy. The adiabatic curves are computed by incorporating explicitly the contribution of molecular vibrational and rotational degrees of freedom. Using a generalized expression for the adiabatic lapse rate, we estimate the adiabatic lapse rate in the troposphere of Titan and Venus. Our results show that the van der Waals-type EOS reproduce observed lapse rates more accurately than the van der Waals EOS.

}




\maketitle

\section{Introduction}\label{sec:intro}

In thermodynamics, an equation of state (EOS) is a fundamental relationship between the thermodynamic variables that describe a specific system, including pressure, volume, and temperature, among others \cite{Callen:450289,10.1093/oso/9780192895547.001.0001,zemansky1997heat}.
For instance, the well-known EOS for an ideal gas reads $pV=nRT$, where $p$, $V$, $n$, and $T$ are the pressure, volume, number of moles (amount of substance), and temperature, respectively. Here, $R$=8.31J/Kmol is the gas constant \cite{Callen:450289,10.1093/oso/9780192895547.001.0001,zemansky1997heat}.

Studying equations of state can reveal interesting properties of matter. In particular, the EOS proposed by van der Waals was the first attempt to describe the behavior of real gases. 
This EOS explicitly takes into account the volume occupied by the molecules and their intermolecular interactions, information encoded in two parameters that can be determined by comparing the EOS with experimental data.
In consequence, the van der Waals EOS adequately predicts the gas-liquid phase transition \cite{Callen:450289}.

One way to determine the free parameters of the van der Waals EOS is to fit the second virial coefficient to the experimental data \cite{wisniak1999interpretation,bugaev2018equation}.
The van der Waals EOS is found to inadequately describe the experimental data at low and high temperatures, especially for polyatomic molecules \cite{bugaev2018equation}. 
Since then, more sophisticated EOS have been developed to accurately reproduce the experimental data, including not only the virial coefficients but also the coexistence curve and other thermodynamic properties of interest.

Among all the proposed modified van der Waals equations of state, Berthelot's equation stands out for its simplicity, as it assumes that molecular attractive interactions vary inversely with temperature \cite{Sobko, WISNIAK2010155}.
In this way, a generalization of the van der Waals EOS consists of considering that the molecular interactions are modeled as a power law in the temperature, which introduces only a single additional parameter\cite{Quintales_1988}. The latter leads to a family of EOS given by
\begin{equation}
    p=\frac{RT}{v-b}-\frac{a}{T^kv^2},
    \label{eq:nvdWEOS}
\end{equation}
where $v=V/n$ is the molar volume and $k$ is a free parameter modeling the explicit dependence of the molecular interaction on the temperature \cite{Quintales_1988}. As usual, the parameters $b$ and $a$ have the same meaning as in the van der Waals EOS. Additionally, all the free parameters are gas-dependent and must be determined by analyzing the experimental data.
Note that the particular cases of $k=0$ or $k=1$ recover the functional form of the van der Waals and Berthelot EOS, respectively. In the following, the family of equations of states governed by Eq.~\eqref{eq:nvdWEOS} will be called van der Waals-type EOS.

A notable application of thermodynamics, especially involving adiabatic curves, lies in describing the thermal properties of planetary atmospheres.
The adiabatic lapse rate model combines the information provided by the adiabatic curves with the hydrostatic equations as an attempt to describe the change in temperature with altitude for different astronomical objects.
In recent papers \cite{AlvarezNavarro, DiazALR2020, DiazALR2022}, the authors have demonstrated that the estimation of the lapse rate can be improved by considering EOS that better reproduce the second and third virial coefficients data. 
Additionally, in some instances, such as the case of Mars and Venus, the inclusion of vibrational degrees of freedom has been shown to play a crucial role in the computation of the adiabatic lapse rate \cite{DiazALR2020}.

In this paper, we aim to estimate the adiabatic lapse rate driven by the corrections provided by the van der Waals-type EOS. 
Our particular interest in this EOS family is because they have a good performance in reproducing the thermodynamic data by only including an additional free parameter compared to the van der Waals EOS.
Within this framework, we estimate the adiabatic lapse rate under the atmospheric conditions of Titan and Venus, which are two cases with extreme atmospheres that emphasize the contributions of molecular interactions and vibrational degrees of freedom to the calculations of the lapse rate.

The rest of this paper is organized as follows: In Sec.~\ref{sec:mvdWEOS}, we briefly review the thermodynamics of the van der Waals-type EOS.
In Sec.~\ref{sec:ALR}, we derive the general formula for the adiabatic lapse rate. Section~\ref{sec:results} contains our estimations of the lapse rate under the atmospheric conditions of Titan and Venus. 
Finally, we wrap up this paper with the final comments and conclusions in Sec.~\ref{sec:conclusions}.

\section{Themodynamics of the van der Waals-type EOS}\label{sec:mvdWEOS}

In this section, we derive key thermodynamic properties of the van der Waals-type EOS introduced in Eq.~\eqref{eq:nvdWEOS}.
It is important to note that Eq.~\eqref{eq:nvdWEOS} can be expressed in terms of reduced variables by determining the critical point, which satisfies the following conditions:
\begin{equation}
    \left( \frac{\partial p}{\partial v} \right)_{T=T_c} = 0
    \quad \mbox{and} \quad
    \left( \frac{\partial^2 p}{\partial v^2} \right)_{T=T_c} = 0, \nonumber
\end{equation}
from which we obtain
\[
v_c=3b, \quad T_c= \left( \frac{8a}{27Rb} \right)^{\frac{1}{k+1}},\quad\mbox{and}\quad p_c=\frac{R}{8b} T_c.
\]
By considering the critical point, Eq.~\eqref{eq:nvdWEOS} can be rewritten in the following form:
\begin{equation}
    p_r=\frac{8T_r}{3v_r-1}-\frac{3}{T_r^k v_r^2},\label{eq:rednvdWEOS}
\end{equation}
where $p_r=p/p_c$, $v_r=v/v_c$, and $T_r=T/T_c$, are the reduced variables.
Additionally, the compressibility factor at the critical point is $Z_c=p_c v_c/R T_c=3/8$, which matches with the estimation for the van der Waals EOS, even when the intermolecular forces differ.

\subsection{Spinodal and coexistence curves}

\begin{figure}
    \centering
    \includegraphics[scale=0.9]{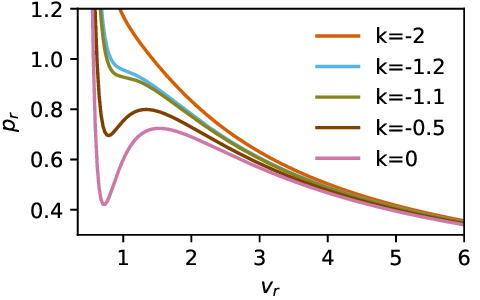}
    \caption{Examples of isotherms ($T_r$=0.9) with different values of $k$. Notice that in the cases where $k<-1$, the isotherms do not exhibit the critical points necessary for the construction of the spinodal curve.}
    \label{fig:isotermask}
\end{figure}

Estimating the spinodal and coexistence curves is essential for identifying the stability and metastability regions.
Since the van der Waals-type EOS is also a cubic EOS, the isotherm curves exhibit two critical points for $T_r<1$, which can be analytically computed by solving the equation $(\partial p_r/\partial v_r)_{T_r}=0$, obtaining the following relation
\begin{equation}
    T_r=\left(\frac{(3v_r-1)^2}{4v_r^3}\right)^{\frac{1}{k+1}}.\label{eq:Trspinodal}
\end{equation}
Substituting \eqref{eq:Trspinodal} into the reduced EOS~\eqref{eq:rednvdWEOS}, we find that the spinal curve takes the following form

\begin{equation}
    p_r = \frac{ 12v_r - 8}{ \left( 4v_r^3(3v_r - 1)^{2k} \right)^{\frac{1}{k+1}} }.
    \label{eq:spinodal}
\end{equation}
Note that the exponent in the denominator of Eq.~\eqref{eq:spinodal} diverges for $k=-1$, which marks a limit for the value of $k$.
In fact, if $k<-1$, the isothermals do not exhibit the two characteristic critical points for $T<T_c$, rendering the model non-physical in this case, as depicted in Fig.~\ref{fig:isotermask}.

\begin{figure}
    \centering
    \includegraphics[scale=0.9]{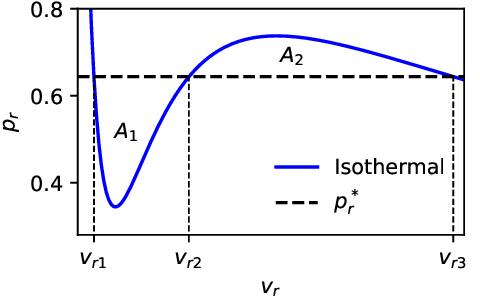}
    \caption{Sketch of the Maxwell construction. $A_1$ and $A_2$ denote the area encompassed between isotherm and the line with constant pressure $p_r^*$. The values $v_{r1}$, $v_{r2}$, and $v_{r3}$ are the points where the isotherm takes the value $p_r^*$, which are found by solving the equation $p_r(v_r)=p_r^*$.}
    \label{fig:maxwell}
\end{figure}

On the other hand, we construct the coexistence curve using the Maxwell construction to address the nonphysical behavior of the isothermal curves for $T<T_c$.
This method \cite{Quintales_1988}, aims to find the value of the pressure at which the areas encompassed between the line $p_r^*$ and the isothermal curve in the regions $v_{r1}$ and $v_{r2}$, and between $v_{r2}$ and $v_{r3}$, take the same value (see Fig.~\ref{fig:maxwell}).
As a first step, we choose the test pressure $p_r^*= \left( p_0+ p_1 \right)/2$, where $p_0=p_r(v_{r\text{min}})$ and $p_1=p_r(v_{r\text{max}})$ are the evaluation of the EOS on the minimum and maximum of the isothermal curve.
The volumes $v_{r1}$, $v_{r2}$, and $v_{r3}$ are the (ordered) solutions of the equation $p_r(v_r)=p_r^*$.
Additionally, the area below and above the line $p_r^*$ and the isotherm are given by
\begin{align}
    A_1&=p_r^* (v_2 - v_1) - \int_{v_1}^{v_2} p_r  dv_r, \\
    A_2&=  \int_{v_2}^{v_3} p_r \, dv_r - p_r^* (v_3 - v_2),
\end{align}
where
\begin{equation}
    \int_{a}^{b} p_r  dv_r=\frac{8 T_r}{3} \ln \left( \frac{3 b - 1}{3 a - 1} \right) + \frac{3}{T_r^k} \left( \frac{1}{b} + \frac{1}{a} \right).
\end{equation}
Within the values of $A_1$ and $A_2$, we define $\Delta A=|A_1-A_2|$ to assess the proximity between the areas $A_1$ and $A_2$.
Since the search for $p_r^*$ is by iterative numerical inspection, we establish the following criterion
$\Delta A<\epsilon$ to determine up to a certain degree of error.
In our computational implementation, we set $\epsilon=0.00001$.
In the case of $\Delta A>\epsilon$, we update the value of $p_r^*$ in the following way: If $A_1-A_2<0$, we replace $p_0$ with $p_r^*$. Otherwise, we set $p_1=p_r^*$.
In both cases, a new value of $p_r^*$ is computed through $p_r^*=\left( p_0+ p_1 \right)/2$.
This process is repeated iteratively until the desired accuracy is achieved.

\begin{figure}
    \centering
    \includegraphics[scale=0.95]{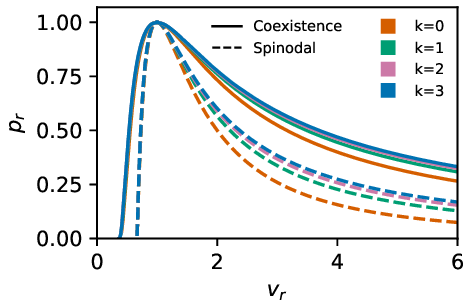}
    \caption{Coexistence (solid lines) and spinodal (dashed lines) curves for the van der Waals-type EOS with different values of $k$. In particular, the cases $k=0$ and $k=1$ correspond to the van der Waals and Berthelot EOS, respectively.}
    \label{fig:coexistence}
\end{figure}

In Fig.~\ref{fig:coexistence} we show our results of the coexistence and spinodal curves.
Note that both curves behave similarly to the van der Waals EOS for $v_r<1$, predicting the same phase transition for $v_r\leq1/3$. 
In contrast, for $v_r>1$, the curves depart from the van der Waals EOS as a consequence of the intermolecular forces' dependence on the temperature.

\subsection{Entropy}

The entropy can be directly derived from the second law of thermodynamics,
\begin{equation}
ds = \frac{1}{T} du + \frac{p}{T} dv
\label{eq:secondlaw},
\end{equation}
where $s=S/n$, $u=U/n$, and $v=V/n$, are molar variables \cite{Callen:450289}.
Additionally, the molar internal energy must satisfy the following relation:
\begin{equation}
\left( \frac{\partial u}{\partial v} \right)_T =T \left( \frac{\partial p}{\partial T} \right)_v - p  =\frac{a(k + 1)}{T^k v^2},
\end{equation}
from which we obtain
\begin{equation}
    u =  f(T) - \frac{a(k + 1)}{T^k v},
    \label{eq:uEOS}
\end{equation}
where $f(T)$ is, in principle, an arbitrary function of temperature, but in Sec.~\ref{sec:adiabaticcurves} we further discuss its relation with the internal degrees of freedom of the molecules, such as vibration and rotation.
Therefore,
\begin{equation}
du=\left( f'(T) +\frac{a(k+1)k}{T^{k+1}v} \right) dT +\frac{a(k+1)}{T^kv^2}dv.
\label{eq:du}
\end{equation}
Substituting \eqref{eq:du} in \eqref{eq:secondlaw} leads to
\begin{equation}
 ds = d \left( F(T)+ \ln(v - b)^R - \frac{ak}{T^{k+1} v} \right),
\end{equation}
where $F(T)$ is an arbitrary function satisfying $F'(T)=f'(T)/T$.
Finally, we identify the entropy as
\begin{equation}
    s = s_0+F(T)+ R\ln(v - b) - \frac{ak}{T^{k+1} v},
\end{equation}
with $s_0$ being a constant associated with the value of the entropy at $T=0$.

\subsection{Adiabatic curves} \label{sec:adiabaticcurves}

In thermodynamics, an adiabatic process corresponds to an idealized thermodynamic process in which the system undergoes no heat exchange with its surroundings. 
Therefore, the internal energy of the system changes solely due to the work done on or by the system.

We derive the adiabatic curve by considering the Helmholtz free energy for an arbitrary EOS, given by \cite{zemansky1997heat, Matsumoto}: 
\begin{equation}
A = -\int p dV + \phi(T)
\label{eq:Helmholtz}
\end{equation}
In Eq.~\eqref{eq:Helmholtz}, the first term corresponds to the mechanical work done by the system during a volume exchange, and $\phi$ is an arbitrary temperature-dependent function.
Additionally, from statistical mechanics, we know that the Helmholtz free energy is related to the partition function of the canonical ensemble through the relation $A=-k_BT\ln Q$ \cite{mcquarrie2000statistical}.
Therefore, for a gas composed of molecules with rotational and vibrational degrees of freedom, we found
\begin{equation}
    \phi(T) = k_B T \ln N! - Nk_B T \ln \left( \frac{q_{\text{tras}}}{V} q_{\text{rot}} q_{\text{vib}} \right),
    \label{eq:phi}
\end{equation}
where
\begin{align}
    \frac{q_{\text{tras}}}{V} =& \left( \frac{2\pi M k_B T}{h^2} \right)^{3/2}\\
    q_{\text{rot}} =& \frac{T^{f_r/2}}{\theta_{\text{rot}}}\\
    q_{\text{vib}} =& 2^{-m}\prod_{j=1}^{m}\text{csch}\left( \frac{h\nu_j}{2k_BT}  \right)
\end{align}
are the translational, rotational, and vibrational contributions to the partition function, respectively.
Additionally, $f_r$ is the number of rotational degrees of freedom, $\theta_\text{rot}$ is a characteristic rotational temperature, and $\nu_j$ are the vibrational frequencies of the molecules.
In this way, the internal energy can be derived from the derivatives of the partition function, yielding
\begin{equation}
U =  \int \left( T \frac{\partial p}{\partial T} - p \right) dV - k_B T^2 \bar{\phi}'(T),
\label{eq:Uadiabatic}
\end{equation}
with $\bar{\phi}(T)=\phi(T)/k_BT$. Note that Eq.~\eqref{eq:Uadiabatic} reproduces Eq.~\eqref{eq:uEOS} for the van der Waals-type EOS, from which we identify $f(T)=- k_B T^2 \bar{\phi}'(T)$, which is explicitly related to the information coming from the molecular degrees of freedom.

On the other hand, the adiabatic curves in the $V$-$T$ representation can be obtained by integrating $dU+pdV=0$, yielding for one mole of gas
\begin{equation}
\int \frac{\partial p}{\partial T} dV - \int \frac{1}{T} \frac{d}{dT} \left( k_B T^2 \bar{\phi}'(T) \right) dT = \text{constant}.
\label{eq:acgeneral}
\end{equation}
Substituting the expression for $\phi(T)$ of Eq.~\eqref{eq:phi} on Eq.~\eqref{eq:acgeneral}, we obtain
\begin{equation}
\int \frac{\partial p}{\partial T} dV + C_V^{IG} \ln T + \frac{R T q_{\text{vib}}'}{q_{\text{vib}}} + R \ln q_{\text{vib}} = \epsilon_0,
\label{eq:acgeneral2}
\end{equation}
where $C_V^{IG}=(3+f_r)R/2$ is the heat capacity of ideal gases and $\epsilon_0$ groups all possible constants.
We emphasize that Eq.~\eqref{eq:acgeneral2} is the most general possible equation in quadratures of the adiabatic curves of one mole of gas. 

In particular, for the van der Waals-type EOS, Equation \eqref{eq:acgeneral2} becomes
\begin{multline}
    R\ln{(V-b)}-\frac{ak}{T^{k+1}V}+ C_V^{IG} \ln T +\\ \frac{R T q_{\text{vib}}'}{q_{\text{vib}}} +
R \ln q_{\text{vib}} = \epsilon_0, \label{eq:adc}
\end{multline}
which has no solution in analytic form, requiring numerical methods to determine the adiabatic volume $V=V(T)$.
In Sec.~\ref{sec:ALR}, we compute the adiabatic curves under the atmospheric conditions of Titan and Venus.

\subsection{Virial coefficients and determination of free parameters}

The free parameters of the van der Waals-type EOS can be determined by comparing the observables derived from the equation of state with experimental data.
For instance, by considering the data of the second viral coefficient.

\begin{figure}
    \includegraphics[scale=0.8]{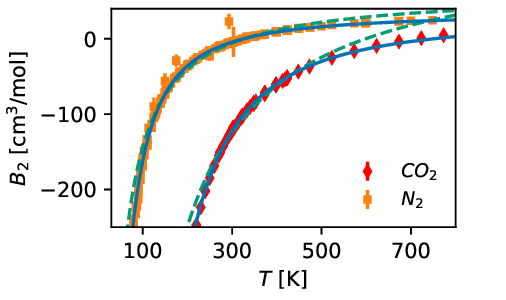}
    \caption{Best fit of second virial coefficient data for N$_2$ (square) and CO$_2$ (diamond). Solid and dashed lines correspond to the second virial function of the van der Waals-type and the van der Waals EOS, respectively. Data taken from Refs.~\cite{845111}.}
    \label{fig:fitB2}
\end{figure}

We use the compressibility factor $Z=pv/RT$ to determine the virial expansion of the van der Waals-type EOS, which reads
\begin{equation}
    Z=\frac{v}{v-b}-\frac{a}{RT^{k+1}v}.
    \label{eq:compressibility}
\end{equation}
Note that the first term of the right hand side of Eq.~\eqref{eq:compressibility} can be rewritten as $v/(v-b)=1/(1-b/v)$. If $b/v<1$, which occurs for the gas phase, the compressibility factor takes the form
\begin{equation}
    Z=\sum_{i=0}^\infty \left( \frac{b}{v}  \right)^i-\frac{a}{RT^{k+1}v}.
    \label{eq:virialexpansion}
\end{equation}
From Eq.~\eqref{eq:virialexpansion}, we identify the second virial coefficient as follows
\begin{equation}
    B_2(T)= b-\frac{a}{RT^{k+1}},
\end{equation}
while the rest of the viral coefficients take the form $B_i=b^i$ for $i>2$.
The free parameters $a$, $b$, and $k$ can be determined by fitting the functional form of $B_2(T)$ to the reported experimental data \cite{Callen:450289}.
Figure~\ref{fig:fitB2} shows the fitted $B_2(T)$ to the experimental data of the second virial coefficient for N$_2$ and CO$_2$. 
Additionally, we report the value of the fitting parameter in Table~\ref{tab:fit_values}.
In both cases, the value of $k$ differs from zero, indicating that the EOS for the molecular nitrogen and carbon dioxide departs from the van der Waals representations. In fact, in the case of carbon dioxide, the value of $k$ is close to one, meaning that the van der Waals-type EOS is closer to the Berthelot EOS than the van der Waals EOS.

\begin{table}
\caption{Fitted values for the parameters $a$, $b$, and $n$ for $N_2$ and $CO_2$.}
\centering

\begin{tabular}{c c c }
\hline
Parameter & CO$_2$ & N$_2$  \\ \hline
    $a$ (m$^6$PaK$^k$/mol$^2$) & $7.0(7) \times 10^7$ & $1.53(4) \times 10^6 $ \\ 
    $b$ (m$^3$/mol) & $26(1) $  & $34.1(3)$ \\ 
    $k$  & 0.92(2) & 0.486(6) \\ \hline
    \end{tabular}%

\label{tab:fit_values}
\end{table}

\section{Adiabatic lapse rate}\label{sec:ALR}

The adiabatic lapse rate is defined as the rate of change of temperature with respect to the variations of altitude for the lowest layers of the atmosphere (troposphere) of an astronomical object (including the Earth), which is given by
\begin{equation}
    \Gamma=-\frac{dT}{dz},
\end{equation}
where $T$ and $z$ are the temperature and altitude, respectively \cite{Jacobson_2005, 10.1093/acprof:oso/9780199562091.001.0001}.

To derive an analytical approach to the adiabatic lapse rate, we start by considering the definition of the compressibility factor $Z=pV/RT$ for a mole of gas, which is a function that depends only on $V$ and $T$.
Additionally, we chose the representation of the adiabatic curves in the $V$-$T$ diagram, as discussed in Sec.~\ref{sec:adiabaticcurves}.
Under these considerations, we have
\begin{equation}
    dp=\frac{R}{V} \left( Z+T\frac{\partial Z}{\partial T} -ZT \frac{V'}{V} +T\frac{\partial Z}{\partial V} V' \right) dT,
    \label{eq:dPZ}
\end{equation}
where $V'$ denotes the derivative of the volume with respect to the temperature.

Furthermore, the hydrostatic equation, $dp=-\rho g dz$, relates the pressure variations with respect to altitude for a gas with density $\rho=M_\text{mol}/V$, being $M_\text{mol}$ the molar mass of the gas in the atmosphere of the astronomical object under study. Thus, the hydrostatic equation takes the following form
\begin{equation}
    dp=-\frac{M_\text{mol}g}{V} dz.
    \label{eq:hydro}
\end{equation}
Therefore, the adiabatic lapse rate is obtained by substituting \eqref{eq:hydro} in \eqref{eq:dPZ}, given by
\begin{equation}
\Gamma=\Gamma^\text{IG}\frac{C_p^\text{IG}}{R}\left( Z+T\frac{\partial Z}{\partial T} -ZT \frac{V'}{V} +T\frac{\partial Z}{\partial V} V'  \right)^{-1}.
\label{eq:lapserate}
\end{equation}
In Eq.~\eqref{eq:lapserate}, $\Gamma^\text{IG}=M_\text{mol}g/C_p^{IG}$ is the lapse rate for the ideal gas model with $C_p^\text{IG}=(5+f_r)R/2$ the heat capacity for the ideal gas \cite{DiazALR2020, DiazALR2022}.

We must emphasize that the adiabatic lapse rate formula in Eq.~\eqref{eq:lapserate} is valid for all EOS and all terms must be evaluated on the adiabatic curves considering the atmospheric conditions of the astronomical objects under study.

\section{Results}\label{sec:results}

In this section, we provide a detailed discussion of the estimations of the adiabatic lapse rate under the atmospheric conditions of Titan and Venus, whose major constituents are N$_2$ (94.2\%) and CO$_2$ (96\%) \cite{CATLING2015429}, respectively.
Regarding the microscopic information required for the calculations, we use the following: due to the linearity of N$_2$ and CO$_2$, the rotational degrees of freedom are $f_r=2$ \cite{10.1093/oso/9780192895547.001.0001}. 
Additionally, the vibrational temperatures are computed through the wave numbers: 
i) 2328.72 cm$^{-1}$ for N$_2$ \cite{dataN2}, and ii) 667.3 cm$^{-1}$ (with degeneracy 2), 1341.5 cm$^{-1}$, and 2349.3 cm$^{-1}$ for CO$_2$ \cite{dataCO2}.
Table~\ref{tab:data} contains all the information needed for the lapse rate computation, including the values corresponding to the ideal gas case ($\Gamma^\text{IG}$) and the estimated value with the data collected by different space missions ($\Gamma^\text{Obs}$) \cite{kasprzak1990pioneer, Lindal1983}.

\begin{table*}[ht]
\caption{Information required for the adiabatic lapse rate estimation of Titan and Venus. Data taken from~\cite{CATLING2015429}. The observed lapse rate for Venus and Titan is reported in Refs.~\cite{Lindal1983, kasprzak1990pioneer}.}
    \centering
    \begin{tabular}{ccccccc}
    \hline
     Astronomical object    & Composition & $\Gamma^\text{IG}$ (K/km) & $\Gamma^\text{Obs}$ (K/km) & $P_0$ (kPa) & $T_0$ (K) & g (m/s$^2$) \\
     \hline
       Titan  & N$_2$ (94.2\%) & 1.30 & 1.38 & 150 & 94 & 1.35 \\
       Venus & CO$_2$ (96.5\%) & 13.42 & 8.40 & 9200 & 737 & 8.87 \\
    \hline
         
    \end{tabular}
    \label{tab:data}
\end{table*}

\begin{figure}
    \includegraphics[scale=0.9]{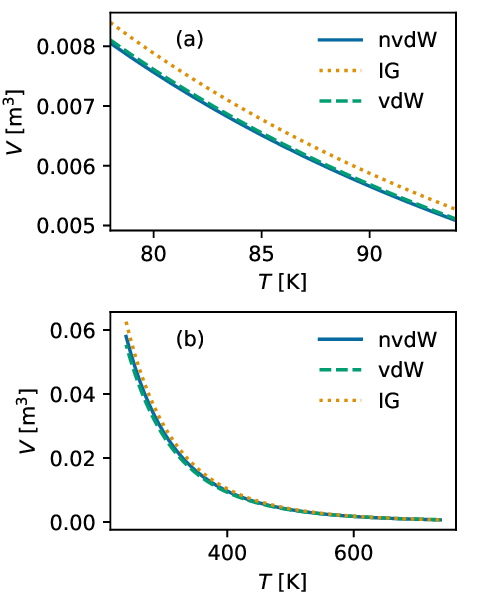}
    \caption{Adiabatic curves of one mole of gas under the atmospheric conditions of (a) Titan and (b) Venus for the ideal gas (dotted line), van der Waals EOS (dashed line), and the van der Waals-type EOS (solid line).}
    \label{fig:adc}
\end{figure}

In all cases, we consider that the atmospheres of Titan and Venus are composed of the major constituents, neglecting the contributions of other gases.
Note that the atmospheric conditions given consider only the information of the temperature and pressure on the surface of the astronomical object, so we compute the corresponding volume $V_0$ by solving the van der Waals-type EOS by considering the values $T_0$ and $P_0$ for one mole of gas. The value of $V_0$ is the initial point from which the adiabatic curve is calculated. 
This is done by numerically solving Eq.~\eqref{eq:adc}.
Figure~\ref{fig:adc} shows our results of the adiabatic curves for the van der Waals-type EOS under the atmospheric conditions of Titan and Venus.
Additionally, we also included the adiabatic curves for the ideal gas model and the van der Waals EOS.
We want to point out that in the case of the van der Waals EOS, the adiabatic curves have an analytic solution, which is obtained directly by solving Eq.~\eqref{eq:acgeneral2} or substituting $k=0$ in Eq.~\eqref{eq:adc}. In all cases, the adiabatic curve deviates from the ideal gas model as the temperature decreases, with the deviation being more notable in the atmospheric conditions of the top troposphere of Venus.

The calculation of the lapse rate through Eq.~\eqref{eq:lapserate} also necessitates the derivatives of the adiabatic curves, which are numerically computed using the five-point stencil method with a step size $\Delta T=0.00001$.
Thus, by plugging in $V$ and $V'$ in Eq.~\eqref{eq:lapserate}, we found the variation of the adiabatic lapse rate with respect to temperature, i.e., $\Gamma=\Gamma(T)$.
However, to estimate the variation of the temperature as a function of the altitude, we need to solve the following differential equation
\begin{equation}
    dz=-\frac{dT}{\Gamma(T)},
\end{equation}
which is performed by using the Runge-Kutta fourth-order method.
In this way, the composition $\Gamma(T(z))$ provides us the description of the adiabatic lapse rate as a function of the altitude, which varies as $z$ runs, in contrast with the ideal gas model which predicts a constant lapse rate value.

\begin{figure}
    \includegraphics[scale=0.9]{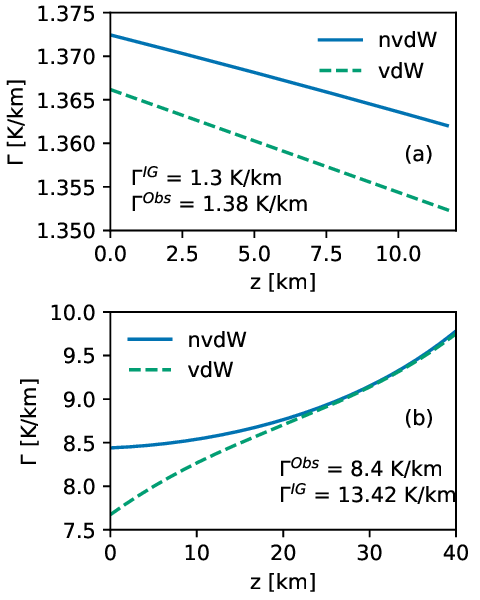}
    \caption{Adiabatic lapse rate estimations for (a) Titan and (b) Venus, considering the van der Waals EOS (dashed line) and the van der Waals-type EOS (solid line).}
    \label{fig:lapse_rate}
\end{figure}

Figure~\ref{fig:lapse_rate} shows our estimations of the adiabatic lapse rate as a function of altitude for Titan and Venus.
Note that the van der Waals-type EOS more accurately reproduces the observed values near the surface of the astronomical objects than the ideal gas model and the van der Waals EOS.
This may occur because the modified EOS provides a better description of the data for the second virial coefficient in the range of temperatures where the lapse rate is estimated.
On the other hand, in the case of the atmospheric conditions of Venus, the estimations provided for both EOS matches at the top of the troposphere.
This last fact is a consequence of atmospheric dilution as altitude increases, where the main contribution to the adiabatic lapse rate comes from molecular vibrations. On the other hand, at the surface level, the large pressure conditions and the molecular interactions become relevant, which produce significant differences between the estimations of the two models.

\section{Conclusions}\label{sec:conclusions}

In this paper, we have revisited the thermodynamic properties of the van der Waals-type EOS, which is a family of equations that generalizes the van der Waals and Berthelot EOS by including a single parameter that controls the temperature dependence of molecular forces in the form of a power law.
As usual, the parameters of this EOS can be found by fitting the data related to the second virial coefficient. 
In the cases of N$_2$ and CO$_2$, we found that the best fit of the EOS parameters indicates that these gases require alternative descriptions than those provided by the van der Waals and Berthelot EOS.
Our interest in N$_2$ and CO$_2$ relies on the fact that they are the principal components of the atmospheres of Titan and Venus, respectively.

In particular, we computed the adiabatic curves for the van der Waals-type EOS, which were used to estimate the adiabatic lapse rate as a function of the altitude for the atmospheres of Titan and Venus.
We found that the van der Waals-type EOS improves the estimation of the adiabatic lapse rate in situations of extreme atmospheric conditions, such as the surface level of Titan and Venus, where the contribution of the molecular interactions is not negligible.
For Venus at the surface level, we also observed a notable deviation between the estimates from the van der Waals and the van der Waals-type EOS, with the latter being close to the values reported in the literature.
However, as altitude increases, the atmosphere becomes more dilute, and both estimates converge, indicating that the main contribution to the lapse rate now comes from internal degrees of freedom.

Finally, we would like to emphasize that the methodology presented in this manuscript can be applied to the estimation of adiabatic lapse rates for other astronomical objects, including exoplanets.

\section*{Acknowledgments}
J.E.Ramírez acknowledges financial support from Secretaría de Ciencia, Humanidades, Tecnología e Innovación-México under a postdoctoral fellowship (grant number 289198).

\bibliography{ref}

\end{document}